\documentclass[dvips]{imsart}

\usepackage{fancyvrb}
\makeatletter\newcommand\code{\bgroup\@makeother\_\@makeother\~\@makeother\$\@codex}
\def\@codex#1{{\normalfont\ttfamily\hyphenchar\font=-1 #1}\egroup}
\makeatother
\newcommand{\proglang}[1]{{\sffamily\bfseries #1}}
\newcommand{\pkg}[1]{{\fontseries{b}\selectfont #1}}
\DefineVerbatimEnvironment{Code}{Verbatim}{}
\DefineVerbatimEnvironment{CodeInput}{Verbatim}{fontshape=sl}
\DefineVerbatimEnvironment{CodeOutput}{Verbatim}{}
\newenvironment{CodeChunk}{}{}
\usepackage{amsmath,amssymb,amsfonts,float,subfig,algorithmic,algorithm,natbib,graphicx}

\begin{document}
\begin{frontmatter}
\title{Iterative Bias Reduction Multivariate Smoothing in \proglang{R}: The \pkg{ibr} Package}
\runtitle{Iterative Bias Reduction}

\begin{aug}
\author{\fnms{Pierre-Andr\'e} \snm{Cornillon}\ead[label=e1]{pierre-andre.cornillon@supagro.inra.fr}},
\author{\fnms{Nicolas} \snm{Hengartner}\ead[label=e2]{nickh@lanl.gov}}
\and
\author{\fnms{Eric} \snm{Matzner-L\o ber}\corref{}\ead[label=e3]{eml@uhb.fr}}
\affiliation{Montpellier SupAgro, University Rennes 2 and Los Alamos National Laboratory}

\address{Address of P-A Cornillon\\
Statistics, IRMAR UMR 6625,\\
Univ. Rennes 2, \\
35043 Rennes, France\\
\printead{e1}}

\address{Address of N. Hengartner\\
Los Alamos National Laboratory,\\
NW, USA\\
\printead{e2}
}

\address{Address of E. Matzner-L\o ber\\
Univ. Rennes, \\
35043 Rennes, France\\
\printead{e3}\\
}

\end{aug}

\begin{abstract}%
In multivariate nonparametric analysis, sparseness of the
  covariates also called curse of dimensionality, forces one to use
  large smoothing parameters. This leads to a biased smoother. Instead
  of focusing on optimally selecting the smoothing parameter, we fix
  it to some reasonably large value to ensure an over-smoothing of the
  data. The resulting base smoother has a small variance but a substantial
  bias. In this paper, we propose an \proglang{R} package named
  \pkg{ibr} to iteratively correct the initial  bias of the (base) estimator by an
  estimate of the bias obtained by smoothing the residuals. After a
  brief description of Iterated Bias Reduction smoothers, we examine the
  base smoothers implemented in the packages: Nadaraya-Watson kernel smoothers 
and thin plate splines smoothers. Then, we  explain the stopping rules available 
in the package and their implementation. Finally we
  illustrate the package on two examples: a toy example in $\mathbb{R}^2$
  and the original Los Angeles ozone dataset.
\end{abstract}
\end{frontmatter}
%% include your article here, just as usual
%% Note that you should use the \pkg{}, \proglang{} and \code{} commands.

\section{Introduction}
Regression is a fundamental data analysis tool for uncovering
functional relationships for the conditional expectation from pairs of
observations $(X_i,Y_i), i=1,\ldots,n$.  Classical linear regression
is the simplest example of this.  More generally, we can let the data
help determine the general form of the relationship by using one of
the numerous non-parametric regression estimators, such as wavelets
based smoothers, kernel smoothers, and splines smoothers
\citep{buja++1989,cleveland+1988,eubank1988,fan+1996,antoniadis+1995,simonoff1996}.
The latter flexible smoothing methods are implemented as \proglang{R}
functions found in numerous contributed packages.  For instance the
package \pkg{wavethresh} \citep{wavethresh} implements a wavelet based
smoother, the package \pkg{lokern} \citep{lokern} provides a kernel
smoothers and the function \code{smooth.spline} calculates a cubic
spline smoother.  When the number of dependent variables $d$ is
greater than 3 or 4, fully non-parametric regression suffers from the
curse of dimensionality, even for moderate sample sizes (say $n$ being
equal to a few hundred).  As a result, application of fully
non-parametric methods are discouraged in dimensions four and higher.
Instead, the statistical literature encourages using constrained
non-parametric regression models (additive models \citep{hastie+1990},
single and multiple index models and projection pursuit models) to
estimate useful approximations of the conditional expectation.  The
latter methods are provided to the \textsf{R} community in the
contributed package \pkg{mgcv} \citep{mgcv} for additive modelling,
function \code{ppr} for projection pursuit and package \pkg{mda}
\citep{mda} for MARS.

Originating from the machine learning community,
the \textit{boosting} algorithm is also another tool for non-parametric regression
\citep[see][and references therein]{friedman2001}. This fairly recent
and very popular method has numerous variations, such as adaboost (the
original method),  logitboost for classification, and the $L_2$ boosting
for regression.  The interesting feature is that it provides a framework for
combining various weak learners (non-parametric smoothers) into a smoother
that is better than any single smoother that it is composed off.  
Packages for $L_2$ boosting are already
available in \textsf{R}: for instance the package \pkg{mboost} \citep{mboost}
allows for $L_2$ boosting for regression problem as well as logistic boosting
for classification.    For multivariate regression, the $L_2$ boosting algorithm 
has been applied to  component-wise additive modelling 
with classical smoother such as smoothing splines
\citep[see][]{buhlmann+2007}.

Linking the $L_2$ boosting algorithm to an iterative bias correction scheme (see for example
\citet{buhlmann+2007} for boosting of smoothing splines and
	\citet{cornillon++2010} for discussions on $L_2$ boosting of
more general smoothers) provides a statistical interpretation of the
$L_2$ boosting algorithm.  This interpretation was alluded to 
in \citet{ridgeway2000}'s discussion of \citet{friedman++2000} paper on
the statistical interpretation of boosting.   The basic idea
of estimating (and correcting for) the bias of a pilot smoother 
goes back to the concept of \textit{twicing} introduced by
\citet{tukey1977}.  The idea of iterating the bias correction was central to 
\textit{adaptive bagging} algorithm of \citet{breiman1999}.  
More details about statistical properties in
univariate or multivariate smoother can be found in
\citet{buhlmann+2003} or \citet{cornillon++2010}. 

This paper focuses on the computational implementation in
\textsf{R} \citep{Rsoft} of the iterated bias correction procedure for
fully multivariate regression smoothers.
We start in Section two by briefly presenting the concept of iterative bias
reduction and recalling its connection to $L_2$ boosting.
The details of our numerical implementation and a review of available options
in our \textbf{R} package \pkg{ibr} are given in Section three.   The last 
section is devoted to examples.

\section{Iterative bias reduction smoothers}
\subsection{Method}
Suppose that the pairs  
$(X_i,Y_i) \in \mathbb{R}^d \times \mathbb{R}$ are related through the non-parametric regression model
\begin{eqnarray} \label{eq:basic.model}
Y_i &=& m(X_i) + \varepsilon_i, \quad i=1,\ldots,n,
\end{eqnarray}
where $m(\cdot)$ is an unknown smooth function, and the disturbances
$\varepsilon_i$ are independent mean zero and variance $\sigma^2$
random variables that are independent of all the covariates.  It is
helpful to rewrite Equation (\ref{eq:basic.model}) in vector form by
setting $Y=(Y_1,\ldots,Y_n)^t$, $m=(m(X_1),\ldots,m(X_n))^t$ and
$\varepsilon=(\varepsilon_1,\ldots,\varepsilon_n)^t$, to get
\begin{eqnarray}
Y &=& m + \varepsilon.   \label{eq:model.vector}
\end{eqnarray}
Linear smoothers estimate the regression function $m$ evaluated at the 
covariates by linear combinations of the responses that can be compactly written as
\begin{equation} \label{eq:smoother.0}
\widehat m_1 = S_\lambda Y,
\end{equation}
where $S_\lambda$ is an $n \times n$ smoothing matrix with smoothing
parameter $\lambda$.  By slight abuse of language, 
we will sometimes refer to the vector of fitted values $\widehat m = \widehat Y =(\widehat
Y_1,\ldots,\widehat Y_n)^t$ as the smooth of $Y$.
Typical smoothers \citep[see for instance][]{hastie+2001} include bin smoothers, spline based smoothers (regression
splines, smoothing splines, and thin-plate splines), kernel based smoothers (Nadaraya-Watson
kernels and local polynomials smoothers), and series based smoothers (Fourier smoothers and wavelet 
smoothers).  In this paper, we focus only on two common types of smoothers:
Nadaraya-Watson kernels (where $\lambda$ is the bandwidth) and
thin-plate splines (where $\lambda$ is the penalty parameter).  Extensions to other
smoothers can easily be achieved by suitably modifying our theoretical results and 
software. 

The linear smoother 
(\ref{eq:smoother.0}) has bias
\begin{eqnarray*} 
%\label{eq:bias.0}
B(\widehat m_1) = E[\widehat m_1|X] - m = (S_\lambda-I)m
\end{eqnarray*}
and variance 
\begin{eqnarray*}
V(\widehat m_1|X) = \left ( S_\lambda S_\lambda^t \right ) \sigma^2.
\end{eqnarray*}
To estimate the bias, observe
that the residuals $R_1=Y-\widehat m_1=(I-S_\lambda)Y$ 
have expected value 
$E[R_1|X] = m - E[\widehat m_1|X] = (I-S_\lambda)m = -B(\widehat m_1)$.
This suggests estimating the bias by smoothing the negative residuals
\begin{eqnarray*}
%\label{eq:bias.1}
\widehat b_1 := -S_\lambda R_1 = -S_\lambda (I-S_\lambda)Y.
\end{eqnarray*}
Recall that, in multivariate non-parametric analysis, sparseness of the
covariates also called curse of dimensionality, forces one to use
large smoothing parameters $\lambda$. This leads to very biased base
smoother $S_\lambda$. Thus the bias correction in multivariate
non-parametric analysis arises as a natural tool to correct classical
smoother $S_\lambda$. If $\lambda$ is large, not all the bias is usually
removed after a the first correction. To remove the remaining bias,
iterations of the bias reduction step have to be performed. For
instance $k-1$ bias reduction step produces the linear smoother at
iteration $k$:
\begin{eqnarray}
\widehat m_k &=& S_\lambda Y + S_\lambda(I-S_\lambda)Y + \cdots +S_\lambda(I-S_\lambda)^{k-1}Y\nonumber\\
&=& (I - (I-S_\lambda)^k)Y.
\label{eq:mk}
\end{eqnarray}
When $d=1$, the sequence of iterated bias corrected smoothers agrees 
with the $L_2$-boosted smoothers without shrinkage.
For $d>1$, the boosting algorithm is applied component-wise
to additive regression models \citep[see][]{buhlmann+2003}.
This results in a sequence of constrained (additive) approximation of
the fully non-parametric regression function $m$.

For thin-plate splines and kernels smoothers (with suitable kernels, such
as as a Gaussian density function),  each iteration of the bias correction
produces a noisier but less biased smoother.  In the 
limit, the sequence of iterative bias corrected smoothers reproduces the 
raw data \citep{cornillon++2010}.   Thus there is a need for good stopping rules 
for the iterative bias correction algorithm.  

\begin{figure}[h] %  figure placement: here, top, bottom, or page
\begin{center}
\includegraphics[width=7cm]{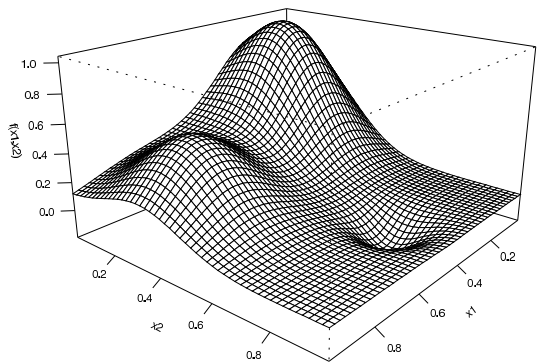}
   \caption{True bivariate regression function $m(x_1,x_2)$ (\ref{eq:mexican.hat})
on the unit square $[0,1]\times[0,1]$ used in our numerical examples.}
   \label{fig:chapeau}
\end{center}
\end{figure}

To illustrate this behavior, let us use a classical
bivariate regression problem: figure \ref{fig:chapeau} graphs
Wendelberger's test function \citep{wendelberger1982}:

\begin{eqnarray}
m(x_1,x_2) &=& \frac{3}{4}\exp \left \{-((9x-2)^2 + (9y-2)^2)/4 \right \} + \nonumber \\
          && \quad +    \frac{3}{4}\exp \left \{-((9x+1)^2/49 + (9y+1)^2/10) \right \} + \nonumber \\
 && \quad       +       \frac{1}{2}\exp \left \{-((9x-7)^2 + (9y-3)^2)/4) \right \} - \nonumber \\
   && \quad -            \frac{1}{5}\exp \left \{-((9x-4)^2 + (9y-7)^2) \right \}. \label{eq:mexican.hat}
\end{eqnarray}

\begin{figure}[h]
\begin{center}
\includegraphics[width=\textwidth]{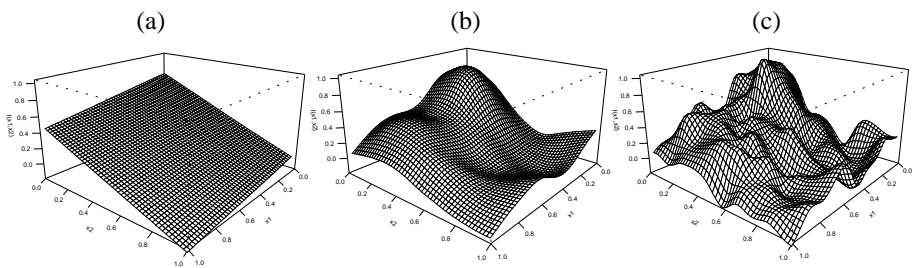}
\caption{Thin-plate spline regression smoothers from $100$ noisy observations from \ref{eq:mexican.hat} (see Figure \ref{fig:chapeau}) computed on a regular 
grid on $[0,1]\times[0,1]$.  Panel (a) shows the pilot smoother, panel (b)
graphs the bias corrected smoother after 500 iterations and panel (c) 
graphs the smoother after 50000 iterations of the bias correction scheme.
\label{fig:exemple0}} 
\end{center}
\end{figure}

The sequence of bias corrected thin-plate spline smoothers, starting
from a pilot that over-smooths the data, converges to an
interpolant of the raw data (see figure
\ref{fig:exemple0} (c)).  After some suitable number of bias
correction steps, the resulting bias corrected smoother will be a good
estimate for the true underlying regression function (see figure
\ref{fig:exemple0} (b)). The crucial choice of $k$ is achieved by the use
of classical criterion such as corrected AIC or GCV (see
section \ref{sec:stoppingrules}).

We note that, provided $\lambda$ is large enough, 
its exact value is not crucial as the
choice of $k$ will adapt to $\lambda$: if two base
smoothers are chosen, one with $\lambda_1$ and another with
$\lambda_2 > \lambda_1$, the chosen iteration $k_2$ will
be greater than $k_1$ as it takes more iterations with a very smooth base
smoother to remove the bias. 

To make prediction at arbitrary locations $x \in \mathbb{R}^d$ 
of the covariates, we extend linear smoothers to functions of the
form
\begin{eqnarray} \label{eq:base.smoother}
\hat m(x) &=& S_\lambda(x)^t Y,
\end{eqnarray}
where $S(x)$ is a vector of size $n$ whose entries are the weights
for predicting $m(x)$.  The vector $S(x)$ reduces to the $j^{th}$ row
of the smoothing matrix when $x=X_j$, and is readily computed for 
many of the smoothers used in practice.    

For extended base smoothers of the form (\ref{eq:base.smoother}), 
we propose to extend the associated iterative bias corrected smoother
$\widehat m_k$ by observing that 
\begin{eqnarray}
\widehat m_k &=& \widehat m_0 + \widehat b_1 + \dots + \widehat b_k \\
&=& S_\lambda[I+(I-S_\lambda)+(I-S_\lambda)^2+\dots+(I-S_\lambda)^{k-1}]Y\label{eq:betak1}\\
&=& S_\lambda \widehat \beta_k\label{eq:betak2}.
\end{eqnarray}
This implies that $\widehat m_k(X_j) = S_\lambda(X_j)^t \hat \beta_k$.
Hence we propose to extend the iterative bias corrected smoother to ${\mathbb R}^d$
via the function
\begin{eqnarray}
\label{entoutx}
\widehat m_k(x) = S_\lambda(x)^t \widehat \beta_k .
\end{eqnarray}
\subsection{Stopping Rules}\label{sec:stoppingrules}
%%%%%%%%%%%%%%%%%%%%%%%%%%%%%% figures
Selecting a suitable number of bias correction iterations $k$ is
crucial. There exists an unknown number $k$ of bias corrections
iterations of thin-plate spline base smoothers that produces
estimators for the regression function that achieve the optimal rate
of convergence for mean square error (see \citep{buhlmann+2003} for
the univariate case and \citep{cornillon++2010} for the multivariate
counter-part).  This optimal unknown number of iterations can be
estimated consistently from data using GCV \citep{cornillon++2010}.

But one can select the number of iterations from data using classical
model selection methodologies such as: GCV \citep{craven1979+}, AIC
\citep{akaike1973}, BIC \citep{sch78}, AICc \citep{hurvich++1998} or
gMDL \citep{hansen+2001}.  In particular, use of AICc is advocated in
\citet{buhlmann+2007} and empirical evidence for GCV can be found in
\citet{cornillon2008++}.  Other methods such as cross-validation
(leave-one-out or $K$-fold) or the use of training set and test set
are also reasonable procedure to estimate $k$ \citep{buhlmann+2007}.
Both empirical \citep{cornillon++2010} and theoretical
\citep{cornillon++2010} considerations support using GCV in practice.

\subsection{Choice of kernel for kernel smoothers}

The behavior of the sequence of iterative bias corrected kernel smoothers
depend critically on the properties of the smoother kernel.  Specifically,
the smoothing kernel needs to be positive definite 
\citep[see][]{marzio+2008,cornillon++2010}.  Examples of positive definite  kernels
include the Gaussian and the triangle densities,
and examples of kernel that are not definite positive include
the uniform and the Epanechnikov kernels.

\section{Implementation in R}

Our implementation of the iterative bias corrected procedure in \proglang{R} follows
the established S3 methods \citep[see writing R extensions][]{Rsoft}.   The main function
(called \code{ibr}) produces an object of class \code{ibr}. 
Applying generic functions, 
such as \code{summary}, \code{predict}, \code{plot} or
\code{residuals}, to an \code{ibr} class object produces 
the expected standard summary statistics, prediction for new data (or fitted values), 
plot of the object and residuals.

\subsection{Base smoother}
Two types of base smoother are implemented in the function \code{ibr}:
thin-plate and kernel smoother. This choice is driven by the
\code{smoother} argument (character): \code{tps} or \code{k}. For
kernel smoother, some classical choice are available using the
\code{kernel} argument (character): Gaussian kernel (\code{g}, the
default), triangle density (\code{t}), and the quartic (\code{q})
density.  The computations have been optimized for the Gaussian
kernel.  Argument \code{kernel} enables the use of Epanechnikov kernel
(\code{e}) or uniform (\code{u}) kernel but only for pedagogical
purposes.

\subsection{Computations}
To predict new data, we compute recursively $\hat \beta_k$ using equations
(\ref{eq:betak1}) and (\ref{eq:betak2}).  Computation of the fitted values
using Equation (\ref{eq:mk}) can be computed using a similar recursive
update formula:   starting with $ \hat b_0=(I-S_\lambda)Y$:
\begin{eqnarray*}
\hat m_k=Y-(I-S_\lambda)\hat b_{k-1}  \ \ \mbox{ and } \ \ 
 \hat b_{k-1}=(I-S_\lambda)b_{k-2}.
\end{eqnarray*}
Computations of either $\hat b_k$ or $\hat \beta_k$ require
$O(kn^2)$ operations and is implemented numerically by using
the corresponding level 2 Blas function \citep{golub+1996}.
In practice, we often found that the number of iterations $k$ that
are required to be evaluated in order to select an good data-driven
choice $\hat k$  is commensurate with the sample size $n$.  Thus
the algorithm that produces the final smoother is typically of order 
$O(n^3)$.

Numerical experiments have shown that an alternative algorithm, 
based on an eigenvalue decomposition of the smoothing matrix $S_\lambda$ 
(also an order $O(n^3)$ algorithm), is faster when combined with GCV for 
selecting the number of iterations.   We have implemented the latter algorithm 
in the \pkg{ibr} package.
This approach is easily understood and implemented for thin plate 
spline smoothers, whose smoothing matrix $S_\lambda$ is symmetric.
For kernel smoothers, the smoothing matrix is not symmetric and 
further discussion is needed.

While the kernel base smoother $S_\lambda$ is not
symmetric, we can rewrite equation (\ref{eq:mk}) using an
eigen decomposition of a symmetric matrix. Specifically, write $S_\lambda= D
\mathbb{K}$, where $\mathbb{K}$ is symmetric matrix with general
element ${\mathbb K}_{ij}
=\prod_{k=1}^dK\left\{(X_{ik}-X_{jk})/h_k\right\}$ and $D$ a diagonal 
matrix with entries $D_{ii}=1/\sum_{j=1}^{n}{\mathbb{K}_{ij}}$. 
With this notation we write the smoothing matrix of $\hat m_k$ in Equation
(\ref{eq:mk}) as
\begin{eqnarray*}
I-(I-S_\lambda)^k&=&I-(I-D\mathbb{K})^k\\
&=&I-(D^{1/2}D^{-1/2}-D^{1/2}D^{1/2}\mathbb{K}D^{1/2}D^{-1/2})^k\\
&=&I-D^{1/2}(I-A)^kD^{-1/2}
\end{eqnarray*}
where $A=D^{1/2}\mathbb{K}D^{1/2}$.  The latter is symmetric, and so can
diagonalized $A=U \Lambda U^\prime$, with $U$ the orthogonal matrix of eigenvectors
and $\Lambda$ the diagonal matrix of eigenvalues.  Equation (\ref{eq:mk})
becomes
\begin{eqnarray*}
\hat m_k&=&D^{1/2}U(I - (I-\Lambda)^k)U'D^{-1/2}Y.
\end{eqnarray*}
The coefficient $\hat \beta_k$  in (\ref{eq:betak1}) becomes
\begin{eqnarray*}
\hat \beta_k &=&D^{1/2}U[I+(I-\Lambda)+(I-\Lambda)^2+\dots+(I-\Lambda)^{k-1}]U'D^{-1/2}Y.
\end{eqnarray*}
Recognizing the sum inside the bracket as the 
$k-1$ first term of geometrical series, we rewrite
\[
\hat \beta_k = D^{1/2}U \Lambda^{-1} (1-(I-\Lambda)^k)U^\prime D^{1/2}.
\]  
 
The core of computation becomes the eigen
decomposition which is done in a very efficient way by the function
\code{eigen} for moderate $n$ ($n<1000$ for instance).  For additional 
efficiencies,  the computations of $A$ and $D^{1/2}$ are done in \proglang{C} 
for the default Gaussian kernel.

\subsection{Stopping rules}
The \pkg{ibr} package implements several 
classical criteria to empirically select an \textit{optimal} 
number  $k$ of bias correction iterations.  They include
Generalized Cross Validation (GCV), the Akaike Information Criteria (AIC), 
the Bayesian Information Criteria (BIC), a corrected Akaike Information Criteria
(AICc) and generalized Minimum Description Length (gMDL). The choice for which
method is to be used is
controlled by the argument \code{criterion} of the \code{ibr} function,
with the default method being GCV. Cross-validation are also available, but
our discussion of that method is postponed to Section \ref{sec:cross-validation}.

The evaluation of an \textit{optimal} number $k$ of iterations using any 
one of these classical criteria
is not a trivial task.  The package \pkg{ibr} implements both a 
computationally burdensome exhaustive search method and a 
computationally efficient but approximate method.  The latter is
the default method.  The user can request \code{ibr} to perform an exhaustive 
search by setting the argument \code{exhaustive=TRUE} in the list
\code{control.par}.

\subsubsection{Exhaustive search method}
The exhaustive search method evaluates,
for each $k$ in an interval $[K_{\min};K_{\max}]$, the criterion 
to identify its global minimizer.  The default values for the range are
$K_{\min}=1$ and $K_{\max}=10^5$.

\subsubsection{Numerical optimization method}
\begin{figure}[H]
\centering
  \subfloat[][Typical evolution of GCV with $k$.\label{fig:gcvtypique}]{\includegraphics[width=0.45\textwidth]{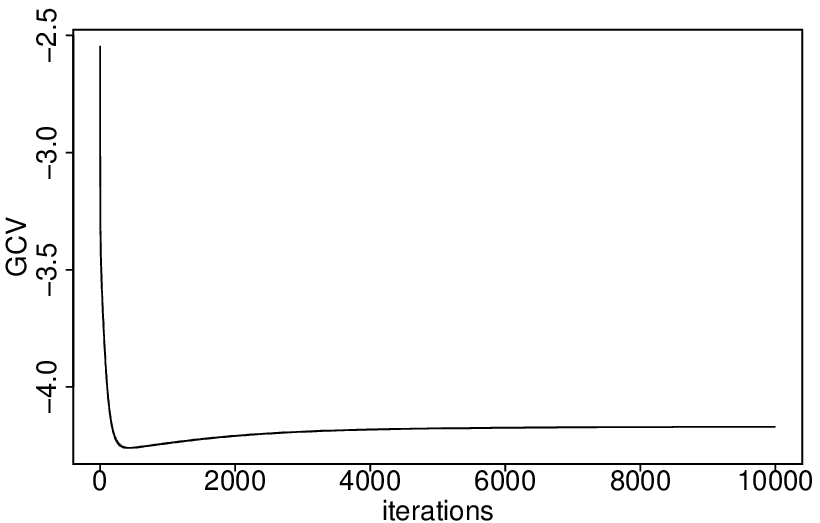}} 
\qquad
  \subfloat[][Evolution of GCV with $k$ with two minima figured by arrows\label{fig:gcv2min}.]{\includegraphics[width=0.45\textwidth]{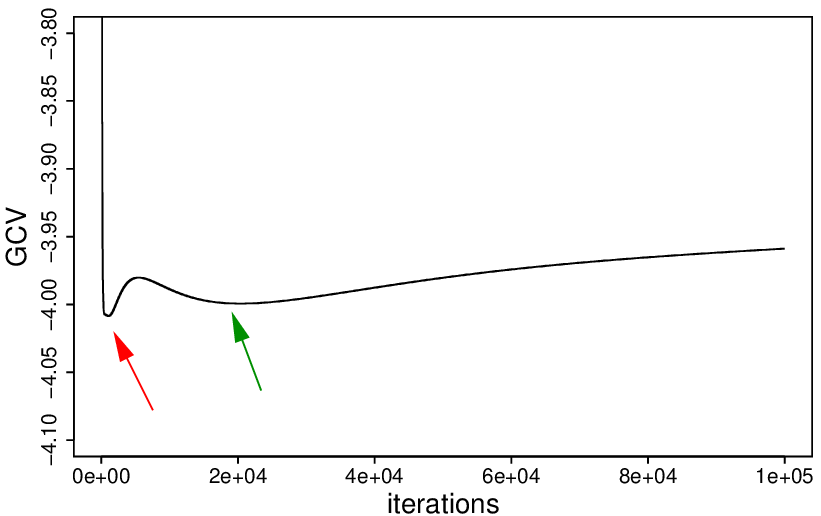}}
\caption{Evolutions of GCV with the number of iterations $k$}
\end{figure}

The default method relies on the fact that the criteria is easily 
calculated for arbitrary $k \in {\mathbb R}^+$.   This 
enables us to use standard optimization routine to minimize
the criterion. While this approach is  conceptually simple,
there are two pitfalls: First, most criterion break down for very 
large  $k$ for which the smoother essentially
interpolates the data, i.e.,  $\hat m_k \approx Y$.   Second, some
criterion exhibit multiple local minima (see figure \ref{fig:gcv2min}).

All model selection criteria trade-off goodness of fit, as measured by 
$\log(\|Y-\hat m_k\|^2)$ with a measure of the complexity of the smoother.
Numerical difficulties arise when $Y \approx \hat m_k$, which occurs when 
the number $k$ of iterations is close to the sample size $n$.  To 
overcome this problem,  we bound from above the maximum allowable
number of iterations by setting the variable  \code{dfmaxi} in the
list \code{control.par}.  By default, its value is $2n/3$.  Hard-coded
error handling prevents evaluation of the criteria when either 
$k > n(1-10^{-10})$ or $\|Y-\hat m_k\|^2 \leq 10^{-10}$.  These 
exceptions also apply to the exhaustive search algorithms.

Classical model selection criteria have been developed in the context
where the effective number of estimated parameters is 
significantly smaller than the number of observations.  Investigation of 
the criteria, as a function of effective degrees of freedom over a broader
range of values reveals the presence of multiple local minima.  While
this does not impact the performance of the exhaustive search, the
presence of local minima is potentially problematic for standard minimization
algorithms.  Our solution is to divide the interval $[K_{\min};K_{\max}]$
into smaller subintervals and apply on each subinterval a numerical 
optimization using the function  \code{optimize}, and the minimizer
of these minimizations is returned.  The splitting is
controlled by the argument \code{fraction} in the list
\code{control.par}, with default value of \code{c(100, 200, 500, 1000,
  5000, 1e04, 5e04, 1e05, 5e05, 1e06)}.

% \begin{figure}[h]
% \begin{center}
% \includegraphics[width=0.5\textwidth]{gcv_2min}
% \caption{Evolution of GCV with $k$ with two minima figured by arrows.}\label{fig:gcv2min} 
% \end{center}
% \end{figure}
While the strategy of optimizing the criteria in subintervals is more expensive
than optimizing over the original interval, it remains significantly faster than
performing an exhaustive search.

\subsection{Scales of variables} \label{section:scale.var} The
function \code{ibr} is designed to be used with two types of linear
smoothers: thin-plate splines and kernel smoothers.
Thin-plate splines are governed by a single parameter $\lambda$ that
weights the contribution of the roughness penalty.  As a result, it is
desirable to scale all the variables to have equal variance to ensure
that the roughness penalty is applied equally to each variable.  This
is achieved by pre-processing the data with the \code{scale} function
before applying smoothing the data with \code{ibr}.

Our implementation of the kernel smoother enables the use of  a vector of different bandwidths, one for each of the regression variables.  While the discussion on scaling
applies when a common bandwidth is used for all the variables, we found
in our numerical experiments that we get better results when we use the 
original variable but select a suitable bandwidth for each variable.
The objective is not to select an optimal bandwidth, but rather control
the amount of smoothing we do at each iteration.  To this end, we propose
to select the bandwidths such that one-dimensional smoothing matrix for 
each variable has the same effective degree of freedom.   Typical values for the
effective degree of freedom values 
are 1.05, 1.1, 1.2, 1.5 or 2, which the user sets with the 
\code{df} argument.  Given an desired effective degree of freedom, 
 package automatically determines the bandwidth using an adaptation 
 of the \code{uniroot} algorithm programed in C.
 
Relating the effective degree of freedom of each of the univariate components to 
an effective degree of freedom for the multivariate smoother is non-trivial.
As a result, some users may prefer to control the overall smoothing
instead of the marginal smoothing of each of component.
To enable (or disable) the control of the overall smoothing, the flag \code{dftotal} in  list \code{control.par} have to be set to \code{TRUE} (the default value of that flag is \code{dftotal=FALSE}).  With
this option, our package takes the value of the argument \code{df}
to calculate the individual bandwidths of each component using 
a \proglang{C} routine. 
 
\subsection{Stopping rules: 
$K$-fold cross-validation and Data splitting}\label{sec:cross-validation}

Simple cross-validation, K-fold cross-validation, and more
generally data splitting, 
are well established techniques for model selection that
we use to determine the optimal number of iteration $k$
for our iterative bias correction scheme.   For these methods, 
the data are separated into two sets, a training set to estimate the 
regression function and a
testing set to evaluate the out of sample prediction error,
using either the root mean square error \code{criterion="rmse"} or the 
mean absolute error \code{criterion="map"} loss functions.   We numerically
minimize that prediction error, either using an optimization routine, the default method,
or by exhaustive search  (set \code{exhaustive=TRUE} in the list \code{control.par}).

\begin{figure}[h]
\begin{center}
\includegraphics{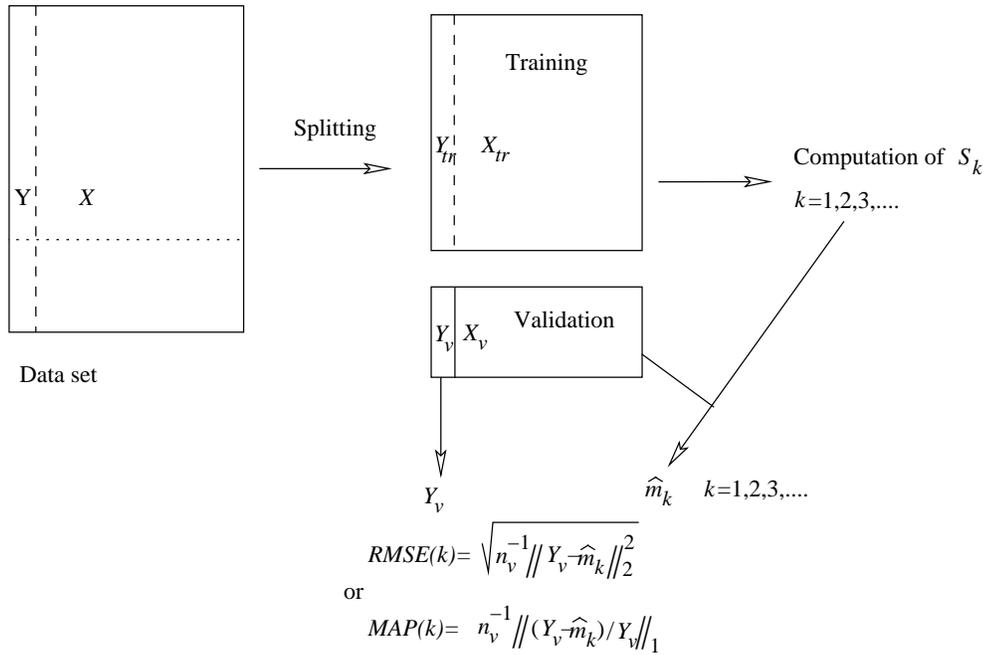}
\caption{Training set and validation set}\label{fig:validation}
\end{center}
\end{figure}

Since simple leave-one out cross-validation usually leads to
estimator that under-smooths (in our case, the selected number of 
iterations $k$ is larger than the optimal one),  we prefer to use either
data splitting or $K$-fold cross-validation. The main difference between
these two procedures is that usually data splitting is conducted
once (except if the user asks for more using argument \code{npermut})
whereas for $K$-fold cross-validation, the original sample is randomly 
partitioned into K subsamples with each of the $K$ subsets used as the test set
and the remainders $K-1$ subsets are combined to form the training set.
The prediction error is then computed by averaging the errors
across the $K$ trials.   In summary, we split
the original data into two samples : a training one on which we evaluate the
estimator and a testing one on which we predict the new observations as shown in
figure \ref{fig:validation}.

The list \code{cv.options} in \code{ibr} controls the various
options for cross-validation, including the size of the training set,
the number of repetition of the procedure, the loss function and
the type of splitting. 

\subsubsection{Selecting the number of iterations $k$ with Data splitting}
To have \code{ibr} perform a data splitting cross-validation,
 we set the following options in
\code{cv.options}:
\begin{enumerate}
\item Input either \code{ntest} or \code{ntrain}, the size of
the testing set $n_v$ or the size of the training test $(n-n_v)$,
respectively.  The default value set \code{ntest} to $\lfloor n/10 \rfloor$.
\item Set the number of times the dataset is split in \code{npermut}.
  For classical data splitting, \code{npermut} have to be set equal to
  one (the default value is 20).
\item Set the \code{type} equal to \code{random} to enable random data
  splitting. This stage can be omitted as this is the default
  value. The argument \code{seed} can be used to control the seed of
  the random generator.
\end{enumerate}
Data splitting (with test set of size $\lfloor n/10 \rfloor$) with root
mean square error loss is achieved by the code
\begin{CodeChunk}
\begin{CodeInput}
> ibr(X,Y,criterion="rmse",cv.options=list(npermut=1))
\end{CodeInput}
\end{CodeChunk}
A more complex example of data splitting that uses 100 samples of 
3 observations to evaluate the prediction error using the mean absolute
deviation loss is achieved with the code
\begin{CodeChunk}
\begin{CodeInput}
> ibr(X,Y,criterion="map",cv.options=list(ntest=3,npermut=100))
\end{CodeInput}
\end{CodeChunk}

\subsubsection{Selecting the number of iterations $k$ with $K$-fold cross-validation}
\begin{figure}[H]
\centering
 \subfloat[][Graphical summary of consecutive $K=6$ folds cross-validation.\label{fig:consecutive}]{\includegraphics[width=0.45\textwidth]{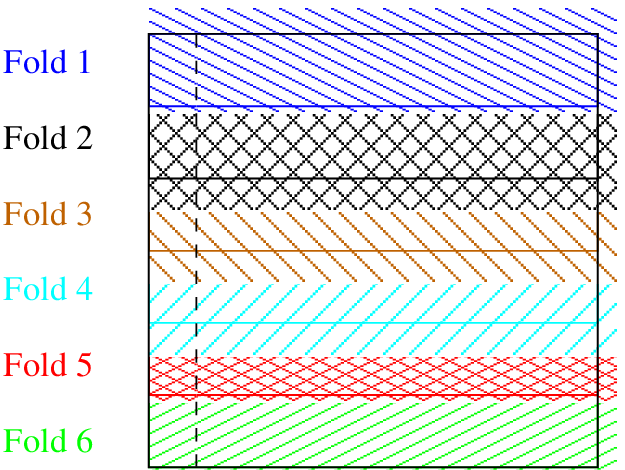}}%
\qquad
 \subfloat[][Graphical summary of interleaved $K=6$ folds cross-validation.\label{fig:interleaved}]{\includegraphics[width=0.45\textwidth]{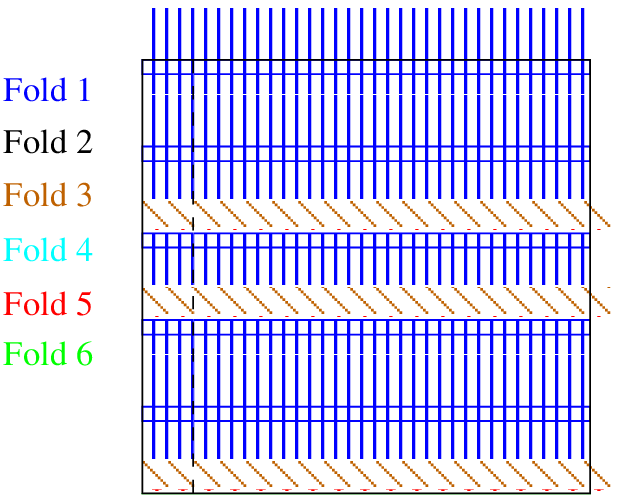}}%
 \caption{Options \code{consecutive} and \code{interleaved} for K-fold
   cross-validation}
\end{figure}

To perform a $K$-cross-validation with \code{ibr}, we 
set the following options in \code{cv.options}:
\begin{enumerate}
\item Set \code{Kfold}=TRUE (default is FALSE) or set \code{Kfold} equal to the number of folds
\item Set the number of folds $K$. One can either specify the size of
  the testing set $n_v$ in \code{ntest} or the size of the training
  set $(n-n_v)$ in \code{ntrain}, in which case the fold is computed
  to be $K=\lfloor n/n_v \rfloor$.  One can set the number of folds
  $K$ using by setting the argument \code{Kfold} equal to $K$. This
  implies that the size of the testing set is $\lfloor n/K \rfloor$.
\item Specify the \code{type} of data-split. By default, the data are
  split randomly (\code{type="random"}).  Alternatively, we divide the
  data using consecutive stretch of data (\code{type="consecutive"}) or
  interleaved split (\code{type="interleaved"}).  A forth option,
  \code{type="timeseries"} divides the data chronologically and uses the
  last $\lfloor n/K \rfloor$ for the testing set.  The splitting implied 
  by \code{consecutive} is shown in figure
  \ref{fig:consecutive} while the splitting using 
  \code{interleaved} is shown in figure \ref{fig:interleaved}. The
  obvious case of random draw is not shown.  Finally, the optional
  argument \code{seed} can be used to control the seed for random
  number generator. It is given as \code{seed} argument of the \code{set.seed}
  function.
\end{enumerate}

The first two lines of code give rise to the examples summarized in
\ref{fig:consecutive} and \ref{fig:interleaved}, while the third line corresponds to
a random K-fold cross-validation:
\begin{CodeChunk}
\begin{CodeInput}
> ibr(X,Y,criterion="rmse",cv.options=list(Kfold=6,type="consecutive"))
> ibr(X,Y,criterion="rmse",cv.options=list(Kfold=6,type="interleaved"))
> ibr(X,Y,criterion="rmse",cv.options=list(Kfold=6,type="random"))
\end{CodeInput}
\end{CodeChunk}

Finally, if the user wants to perform an exhaustive search for the
number of iterations (from 1 to 1000 iterations) using the
\emph{leave-one out} cross-validation, she runs

\begin{CodeChunk}
\begin{CodeInput}
> ibr(X,Y,criterion="rmse",Kmax=1000,control.par=list(exhaustive=TRUE),
+  cv.options=list(Kfold=TRUE,ntest=1,type="consecutive"))
\end{CodeInput}
\end{CodeChunk} 
 
\subsection{Variables selection}\label{sec:selvar}

We can apply the standard strategy of balancing prediction errors and model
complexity to select predictors.  The main issue with variable selection with 
\proglang{ibr} is computational, as we wish to compare models using an optimal number
of bias reduction iterations.  To limit fitting models with many parameters,
we only consider forward variable selection (see algorithm \ref{algoforward}).

In analogy to selecting the number of iterations, controlled by entries
in the list \code{criterion}, we control the variable selection procedure
with the list \code{varcrit}.  The latter has the same default values 
as the former.

\begin{algorithm}[H]
\caption{Forward function}\label{algoforward}
\begin{algorithmic}
\REQUIRE \verb@criterion@ (GCV, AIC, AICc, BIC, gMDL, MAP or RMSE) 
\REQUIRE \verb@varcrit@ (GCV, AIC, AICc, BIC, gMDL) 
\STATE $s\gets 1$ \verb@#@ current stage
\STATE $R$ matrix of infinity with $d$ columns \verb@#@ Matrix of results
\STATE $\mathcal{S}\gets \emptyset$ \verb@#@ variable(s) selected at current stage
\STATE $s_{\min}\gets \infty$ \verb@#@ Current minimum of criterion

\FOR{$s=1$ to $d$} 
  \FOR{$j=1$ to $d$ such that $j\not\in \mathcal{S}$} 
   \STATE $\mathcal{S}_c\gets \mathcal{S}\cup \{j\}$ 
\verb@#@ Adding one variable to the set of variables already selected
   \STATE \verb@res <- ibr(@$X_{\mathcal{S}_c}$\verb@,Y,criterion)@ \verb@#@ 
$X_{\mathcal{S}_c}$ is the dataset with explanatory variables in $\mathcal{S}_c$
   \STATE evaluation of criterion \verb@varcrit@ for \verb@res@: $R_{sj}$     
  \ENDFOR
\IF{all $\left\{R_{sj}\right\}_j > s_{\min}$} 
\STATE \textbf{Return} matrix $R$ from row 1 to $s-1$
\ELSE
\STATE \verb@#@ Updating
\STATE $\mathcal{S}\gets\mathcal{S}\cup \{\arg\min_j R_{sj}\}$
\STATE $s_{\min}\gets \min_j R_{sj}$ \verb@#@ Current minimum of criterion \verb@varcrit@ 
\STATE $s\gets s+1$
\ENDIF
\ENDFOR
\end{algorithmic}
\end{algorithm}

The \code{forward} function returns an object of class
\code{forwardibr}. A plot method is provided for this
class of object.

\section{Examples}
Let us return to the Wendelberger's test function (see equation \eqref{eq:mexican.hat}):
\begin{CodeChunk}
\begin{CodeInput}
>  f <- function(x, y) { .75*exp(-((9*x-2)^2 + (9*y-2)^2)/4) +
+  .75*exp(-((9*x+1)^2/49 + (9*y+1)^2/10)) +
+  .50*exp(-((9*x-7)^2 + (9*y-3)^2)/4) -
+  .20*exp(-((9*x-4)^2 + (9*y-7)^2)) }
\end{CodeInput}
\end{CodeChunk}
We start by plotting this function on a $50\times 50$ grid of points in the unit square
$(0,1) \times (0,1)$ that produces Figure \ref{fig:chapeau}.
\begin{CodeChunk}
\begin{CodeInput}
> ngrid <- 50; xf <- seq(0,1, length=ngrid+2)[-c(1,ngrid+2)]
> yf <- xf ; zf <- outer(xf, yf, f)
> grid <- cbind(rep(xf, ngrid), rep(xf, rep(ngrid, ngrid)))
> persp(xf, yf, zf, theta=130, phi=20, expand=0.45,main="True Function")
\end{CodeInput}
\end{CodeChunk}
Next, we can generate a dataset of 100 noisy observations of the function $f$ evaluated
on the regular grid
$\{0.05,0.15,\ldots,0.85,0.95\}^2$, with Gaussian  disturbances that 
have zero mean and standard deviation producing a signal 
to noise ratio of five.
\begin{CodeChunk}
\begin{CodeInput}
> noise <- .2 ; N <- 100 
> xr <- seq(0.05,0.95,by=0.1) ; yr <- xr ; zr <- outer(xr,yr,f) ; set.seed(25)
> std <- sqrt(noise*var(as.vector(zr))) ; noise <- rnorm(length(zr),0,std)
> Z <- zr + matrix(noise,sqrt(N),sqrt(N))
\end{CodeInput}
\end{CodeChunk}
Concatenate the explanatory variables into a $100 \times 2$ matrix that results in the 
objects \code{X} and \code{Zc}. 
\begin{CodeChunk}
\begin{CodeInput}
> xc <- rep(xr, sqrt(N)) ; yc <- rep(yr, rep(sqrt(N),sqrt(N)))
> X <- cbind(xc, yc) ; Zc <- as.vector(Z)
\end{CodeInput}
\end{CodeChunk}

In this example, we will use thin-plate splines of order $\nu_0$.   Since the procedure
is adaptive, the default value is the smallest possible smoothness, which is $2$ in our
case.  The effective degree of freedom of the thin plate smoother needs to be
slightly larger than  $M_0={\nu_0+d-1\choose \nu_0-1}$.  In our example, $M=3$
and we chose $\lambda$ such that the effective degree of freedom was 
$1.1 \times M = 3.3$.  Figure \ref{fig:exemple0} (a) graphs the base smoother
at iteration zero.

\begin{CodeChunk}
\begin{CodeInput}
> res.ibr <- ibr(X,Zc,df=1.1,control.par=list(iter=1),smoother="tps")
> fit <- matrix(predict(res.ibr,grid),ngrid,ngrid)
> persp(xf, yf, fit ,theta=130,phi=20,expand=0.45,main="Fit",zlab="fit")
\end{CodeInput}
\end{CodeChunk}
Figure \ref{fig:exemple0} (b) and (c) show the bias corrected 
smoother after $500$ and $50,000$ iterations.   To compute the smoother 
whose number of iterations is selected with GCV, we use
\begin{CodeChunk}
\begin{CodeInput}
> res.ibr <- ibr(X,Zc,df=1.1,smoother="tps")
> summary(res.ibr)
\end{CodeInput}
\end{CodeChunk}
The summary output of the resulting smoother prints 
the residuals standard error, the degree initial freedom and reveals that
the final degree of freedom is 26.5 and the value of (log) GCV is 
-3.63 after  iterations $\hat k_{GCV}=424$ iterations.

\begin{CodeOutput}
Residuals:
      Min        1Q    Median        3Q       Max 
-0.235036 -0.068252 -0.007412  0.069061  0.301478 
Residual standard error: 0.1197 on 73.5 degrees of freedom

Initial df: 3.3 ; Final df: 26.5 
  gcv 
-3.63 

Number of iterations: 424 chosen by gcv 
Base smoother: Thin plate spline of order 2 (with 3.3 df)
\end{CodeOutput}
To compute the fitted values, we use the predict function 
\begin{CodeInput}
> predict(res.ibr)
\end{CodeInput}
that can be used to compute the Mean Absolute Error (MAE) on a grid
\begin{CodeInput}
> mean(abs(predict(res.ibr,grid)-as.vector(zf)))
[1] 0.0578394
\end{CodeInput}
To plot the fitted value, we employ the code
\begin{CodeInput}
> predgrid <- matrix(predict(res.ibr,grid),ngrid,ngrid)
> persp(xf,yf,predgrid,theta=130,phi=20,expand=0.45,zlab="fit")
\end{CodeInput}
\begin{figure}[H] %  figure placement: here, top, bottom, or page
   \centering
   \includegraphics[width=4cm]{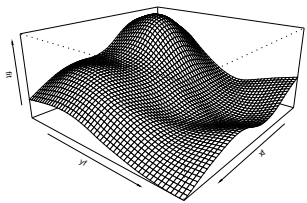} 
   \caption{Fitted regression function $\hat m_k(x_1,x_2)$ on the unit square $[0,1]\times[0,1]$, the number of iteration is chosen by GCV: $\hat k_{GCV}=424$.}
   \label{fig:chapeaufit}
\end{figure}
To use either the  AICc or the BIC criterion to select the number of iterations, we write
\begin{CodeChunk}
\begin{CodeInput}
> res.ibr.aicc <- ibr(X,Zc,df=1.1,smoother="tps",crit="aicc")
> res.ibr.bic <- ibr(X,Zc,df=1.1,smoother="tps",crit="bic")
\end{CodeInput}
\end{CodeChunk}
Direct display of an \code{ibr} object gives the following short description
\begin{CodeChunk}
\begin{CodeInput}
> res.ibr.aicc
Initial df: 3.3 ; Final df: 20.98 
Number of iterations: 247 chosen by aicc 

> mean(abs(predict(res.ibr.aicc,grid)-as.vector(zf)))
[1] 0.0583159
\end{CodeInput}
\end{CodeChunk}
which reveals that AICc required 247 iterations and the resulting smoother
has a slightly larger than mean absolute error then what we obtained using GCV.
This last MAE is close to the thin plates spline smoother with
$\lambda$ (not $k$) selected with GCV 
\begin{CodeChunk}
\begin{CodeInput}
> res.tps <- Tps(X,Zc)
> mean(abs(predict(res.tps,grid)-as.vector(zf)))
[1] 0.05823783
\end{CodeInput}
\end{CodeChunk}
\subsection{Real example: Los Angeles Ozone Data}
\label{section:exampleozone}
We consider the classical Los Angeles basin ozone concentration data set 
used by numerous authors (see for example
\cite{breiman1996,buhlmann+2003,buhlmann+2006}) to demonstrate the 
performance of
various high dimensional smoothing techniques.  The data consists of 
$n=330$ observed ozone concentration related to $d=8$ explanatory
variables. 

The order $\nu_0$ of thin plate splines needs to be greater than $d/2$, 
that is $\nu_0=5$.  This implies that the minimal effective degree of freedom 
of the thin plate spline smoother $S_\lambda$ is $M_0=495$, which is greater than 
the sample size $n$.    Even for larger sample sizes, say $n=500$,   the thin
plate splines will be unsatisfactory base smoother (recall that in the
preceding section, for $d=2$ we started at $3.3$ df with 100
observations).

For this reason, let us consider the (default) Gaussian kernel
smoother.  As we discussed in Section \ref{section:scale.var}, we do
not scale the eight explanatory variables but instead select the
bandwidth of each univariate smoother to achieve a smoothing matrix
that has an effective degree of freedom of $1.1$.  This ensures that
at face value, each of the eight covariates has the same influence.
The number of possible bias correction iterations $k$ considered by
the model selection procedure for selecting the optimal number of
iterations lies between one and 10, 000 (default values for
\code{Kmin} and \code{Kmax}).  The \code{R} code for fitting this data
is
\begin{CodeInput}
> data(ozone)
> res.ibr <- ibr(ozone[,-1],ozone[,1],df=1.1)
> summary(res.ibr)

Residuals:
     Min       1Q   Median       3Q      Max
-13.5581  -2.0566  -0.3481   1.9816  12.6049
Residual standard error: 3.946 on 309.6 degrees of freedom

Initial df: 2.06 ; Final df: 20.42
  gcv
2.873

Number of iterations: 64 chosen by gcv
Base smoother: gaussian kernel (with 2.06 df)
\end{CodeInput}
From the summary, we see that the optimal
number of iterations is $\hat k_{GCV}=64$, which can be thought as
quite low (recall that in the previous example the number of
iterations ranged between $200$ and $400$).   In this example,  
an exhaustive search method for determining the optimal number of 
iterations 
\begin{CodeInput}
> ibr(ozone[,-1],ozone[,1],df=1.1,control.par=list(exhaustive=TRUE))
\end{CodeInput}
gives the same result.   Because we only need a relatively small number
of bias correction steps, 
we can select a smaller initial effective degree of freedom, say
$1.05$, while maintaining the computational complexity at a manageable
level.  Indeed, decreasing the effective degree of freedom of the pilot
smoother increases the total number of bias reduction steps while typically
providing some performance gains as measured by out of sample 
prediction errors.

A plot method is also available for the \code{ibr} object to display 
the residuals as a function of the index. 
\begin{CodeInput}
> plot(res.ibr)
\end{CodeInput}
\begin{figure}[H] %  figure placement: here, top, bottom, or page
   \centering
   \includegraphics{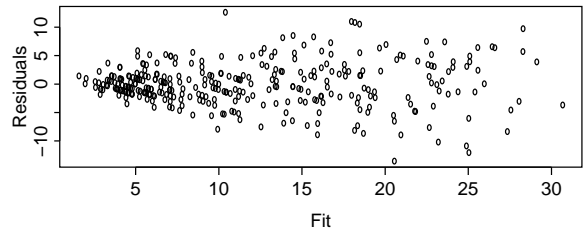}
   \caption{Index plot of residuals.}
   \label{fig:plotibrs}
\end{figure}
To emulate \citep[see][]{buhlmann+2003}  and draw $50$ random splits of the
data into a set of $297$ training data and a set of $33$ testing data, we issue the
following commands
\begin{CodeInput}
> XX <- ozone[,-1]
> Y <- ozone[,1]
> erreur1.5 <- rep(0,33*50)
> aa <- c(1,945095059,162152953)
> for(i in 1:50){
+    set.seed(aa+i)
+    ind <- sample(1:330,33)
+    XXA <- XX[-ind,]
+    YA <- Y[-ind]
+    XXT <- XX[ind,]
+    YT <- Y[ind]
+    res.ibr <- ibr(XXA,YA)
+    erreur1.5[(33*(i-1)+1):(33*i)] <- YT-predict(res.ibr,XXT)
+  }
> print(mean(erreur1.5^2))
\end{CodeInput}
We get an error of $14.98$, which compare favorably with GAM
(\pkg{mgcv}: 17.44), MARS (\pkg{mda}: 17.49), projection pursuit
(\code{ppr}: 17.79 for \code{nterms=2}) or boosting (package
\pkg{mboost}: 17.23).  Note that since no default is available for the \code{nterms}
argument of the function \code{ppr}, we follow the examples provided in the
\code{ppr} documentation and have set  \code{nterms} equal
to 2.  To summarize, the \pkg{ibr} smoother enjoys a 15\% reduction in the 
out of sample prediction mean squared error over other state-of-the-art multivariate
smoothing methods.

We note that the above comparison favors the $L_2$ boosting and MARS 
algorithms that take advantage of built-in variable selection procedures.
To compare to these methods, 
we apply the forward variable selection using the random splitting 
method to \code{ibr} for this data set
issuing the following commands.
\begin{CodeInput}
> aa <- c(1,945095059,162152953)
> i <- 1
> set.seed(aa+i)
> ind <- sample(1:330,33)
> XXA <- XX[-ind,]
> YA <- Y[-ind]
> XXT <- XX[ind,]
> YT <- Y[ind]
\end{CodeInput}
We select variables using the commands 
\begin{CodeInput}
> forward.ibr <- forward(XXA,YA)
> varnumber <- apply(forward.ibr,1,which.min)
> varnumber
[1] 4 3 7 6 5
\end{CodeInput}
That is, the order of the variables to be included into the model is
4, 3, 7, 6 and 5.   Variable selection leads to improved predictions.
To quantify, on the testing set, this improvement, we compare the
the prediction MSE of the selected five variable model with the 
prediction MSE of model that uses all the eight variables.
\begin{CodeInput}
> res.ibr <- ibr(XXA,YA)
> mean((YT-predict(res.ibr,XXT))^2)
[1] 22.90836
> res.ibr2 <- ibr(XXA[,varnumber],YA)
> mean((YT-predict(res.ibr2,XXT[,varnumber]))^2)
[1] 21.12792
\end{CodeInput}
This shows a small improvement.  In
conclusion for this example, we remark that despite the increased
computational time, the \code{forward} function provides simple and useful
tool for selecting variables.

\section{Conclusion}
The \pkg{ibr} package provides additional features which are not
offered by other packages on CRAN. These features are a complete
implementation, using \proglang{R} language, of iterative biased
reduction procedure which implement and generalize the twicing idea of
\citet{tukey1977}.

This method of smoothing for multivariate dataset seems to be promising
especially on real dataset. But one limitation of this smoothing
method is the use of matrix $n\times n$, where $n$ is the number of
observations. Moreover, at the present time, the computational bottleneck 
is the eigen decomposition of an $n\times n$ matrix, which limits the size
of the dataset to which this procedure can be applied too.

%%%%%%%%%%%%%%%%%%%%%%%%%%%%%%%%%%%%%%%%%%%%%%%%%%%%%%%%%%%%%%%%%
\bibliographystyle{abbrvnat} 
\bibliography{./biblio}

\begin{thebibliography}{33}
\providecommand{\natexlab}[1]{#1}
\providecommand{\url}[1]{\texttt{#1}}
\expandafter\ifx\csname urlstyle\endcsname\relax
  \providecommand{\doi}[1]{doi: #1}\else
  \providecommand{\doi}{doi: \begingroup \urlstyle{rm}\Url}\fi

\bibitem[Akaike(1973)]{akaike1973}
H.~Akaike.
\newblock Information theory and an extension of the maximum likelihood
  principle.
\newblock In B.~N. Petrov and B.~F. Csaki, editors, \emph{Second international
  symposium on information theory}, pages 267--281, Budapest, 1973. Academiai
  Kiado.

\bibitem[Antoniadis and Oppenheim(1995)]{antoniadis+1995}
A.~Antoniadis and G.~Oppenheim.
\newblock \emph{Wavelets in Statistics}.
\newblock Lecture Notes in Statistics, Springer Verlag, 1995.

\bibitem[Breiman(1996)]{breiman1996}
L.~Breiman.
\newblock Bagging predictors.
\newblock \emph{Machine Learning}, 24:\penalty0 123--140, 1996.

\bibitem[Breiman(1999)]{breiman1999}
L.~Breiman.
\newblock Using adaptive bagging to debias regressions.
\newblock Technical Report 547, Department of Statistics, UC Berkeley, 1999.

\bibitem[B\"uhlmann and Hothorn(2007)]{buhlmann+2007}
P.~B\"uhlmann and T.~Hothorn.
\newblock Boosting algorithms: regularization, prediction and model fitting
  (with discussion).
\newblock \emph{Statistical Science}, 22:\penalty0 477--505, 2007.

\bibitem[B\"uhlmann and Yu(2003)]{buhlmann+2003}
P.~B\"uhlmann and B.~Yu.
\newblock Boosting with the $l_2$ loss: Regression and classification.
\newblock \emph{J. Amer. Statist. Assoc.}, 98:\penalty0 324--339, 2003.

\bibitem[B\"uhlmann and Yu(2006)]{buhlmann+2006}
P.~B\"uhlmann and B.~Yu.
\newblock Sparse boosting.
\newblock \emph{J. Machine Learning Research}, 7:\penalty0 1001--1024, 2006.

\bibitem[Buja et~al.(1989)Buja, Hastie, and Tibshirani]{buja++1989}
A.~Buja, T.~Hastie, and R.~Tibshirani.
\newblock Linear smoothers and additive models.
\newblock \emph{Ann. of Statist.}, 17:\penalty0 453--510, 1989.

\bibitem[Cleveland and Devlin(1988)]{cleveland+1988}
W.~Cleveland and S.~Devlin.
\newblock Locally weighted regression : an approach to regression analysis by
  local fitting.
\newblock \emph{J. Amer. Stat. Ass.}, 83:\penalty0 596--610, 1988.

\bibitem[Cornillon et~al.(2008)Cornillon, Hengartner, and
  Matzner-L{\o}ber]{cornillon2008++}
P.~A. Cornillon, N.~Hengartner, and E.~Matzner-L{\o}ber.
\newblock Recursive bias estimation and $l_2$ boosting.
\newblock Technical report, arXiv, 2008.

\bibitem[Cornillon et~al.(2011)Cornillon, Hengartner, and
  Matzner-L{\o}ber]{cornillon++2010}
P.~A. Cornillon, N.~Hengartner, and E.~Matzner-L{\o}ber.
\newblock Recursive bias estimation for multivariate regression smoothers.
\newblock Technical report, arXiv, 2011.

\bibitem[Craven and Wahba(1979)]{craven1979+}
P.~Craven and G.~Wahba.
\newblock Smoothing noisy data with spline functions: Estimating the correct
  degree of smoothing by the method of generalized cross-validation.
\newblock \emph{Numerical Mathematics}, 31:\penalty0 377--403, 1979.

\bibitem[Di~Marzio and Taylor(2008)]{marzio+2008}
M.~Di~Marzio and C.~Taylor.
\newblock On boosting kernel regression.
\newblock \emph{to appear in JSPI}, 2008.

\bibitem[Eubank(1988)]{eubank1988}
R.~Eubank.
\newblock \emph{Spline Smoothing and Nonparametric Regression}.
\newblock Marcel Dekker, New-York, 1988.

\bibitem[Fan and Gijbels(1996)]{fan+1996}
J.~Fan and I.~Gijbels.
\newblock \emph{Local Polynomial Modeling and Its Application, Theory and
  Methodologies}.
\newblock Chapman et Hall, New York, 1996.

\bibitem[for R and enhanced~by Martin~Maechler(2010)]{lokern}
E.~H.~P. for R and enhanced~by Martin~Maechler.
\newblock \emph{\pkg{lokern}: Kernel Regression Smoothing with Local or Global
  Plug-in Bandwidth}, 2010.
\newblock URL \url{http://CRAN.R-project.org/package=lokern}.
\newblock R package version 1.1-2.

\bibitem[Friedman(2001)]{friedman2001}
J.~Friedman.
\newblock Greedy function approximation: A gradient boosting machine.
\newblock \emph{Ann. Statist.}, 28\penalty0 (337-407), 2001.

\bibitem[Friedman et~al.(2000)Friedman, Hastie, and Tibshirani]{friedman++2000}
J.~Friedman, T.~Hastie, and R.~Tibshirani.
\newblock Additive logistic regression: a statistical view of boosting.
\newblock \emph{Ann. of Statist.}, 28:\penalty0 337--407, 2000.

\bibitem[Golub and {V}an Loan(1996)]{golub+1996}
G.~H. Golub and C.~F. {V}an Loan.
\newblock \emph{Matrix computations}.
\newblock The Johns Hopkins University Press, 3 edition, 1996.

\bibitem[Hansen and Yu(2001)]{hansen+2001}
M.~Hansen and B.~Yu.
\newblock Model selection and minimal description length principle.
\newblock \emph{J. Amer. Statist. Assoc.}, 96:\penalty0 746--774, 2001.

\bibitem[Hastie et~al.(2011)Hastie, Tibshirani, Leisch, Hornik, and
  Ripley]{mda}
T.~Hastie, R.~Tibshirani, F.~Leisch, K.~Hornik, and B.~D. Ripley.
\newblock \emph{\pkg{mda}: Mixture and flexible discriminant analysis}, 2011.
\newblock URL \url{http://CRAN.R-project.org/package=mda}.
\newblock R package version 0.4-2.

\bibitem[Hastie and Tibshirani(1990)]{hastie+1990}
T.~J. Hastie and R.~J. Tibshirani.
\newblock \emph{Generalized Additive Models}.
\newblock Chapman \& Hall, London, 1990.

\bibitem[Hastie et~al.(2001)Hastie, Tibshirani, and Friedman]{hastie+2001}
T.~J. Hastie, R.~J. Tibshirani, and J.~H. Friedman.
\newblock \emph{The elements of statistical learning: data mining, inference
  and prediction}.
\newblock Springer, New-York, 2001.

\bibitem[Hothorn et~al.(2010)Hothorn, Buehlmann, Kneib, Schmid, and
  Hofner]{mboost}
T.~Hothorn, P.~Buehlmann, T.~Kneib, M.~Schmid, and B.~Hofner.
\newblock \emph{Model-Based Boosting}, 2010.
\newblock URL \url{http://CRAN.R-project.org/package=mboost}.
\newblock R package version 2.0-9.

\bibitem[Hurvich et~al.(1998)Hurvich, Simonoff, and Tsai]{hurvich++1998}
C.~Hurvich, G.~Simonoff, and C.~L. Tsai.
\newblock Smoothing parameter selection in nonparametric regression using and
  improved akaike information criterion.
\newblock \emph{J. R. Statist. Soc. B}, 60:\penalty0 271--294, 1998.

\bibitem[Nason(2010)]{wavethresh}
G.~Nason.
\newblock \emph{\pkg{wavethresh}: Wavelets statistics and transforms.}, 2010.
\newblock URL \url{http://CRAN.R-project.org/package=wavethresh}.
\newblock R package version 4.5.

\bibitem[{R Development Core Team}(2009)]{Rsoft}
{R Development Core Team}.
\newblock \emph{R: A Language and Environment for Statistical Computing}.
\newblock Vienna, Austria, 2009.
\newblock URL \url{http://www.R-project.org/}.

\bibitem[Ridgeway(2000)]{ridgeway2000}
G.~Ridgeway.
\newblock Additive logistic regression: a statistical view of boosting:
  Discussion.
\newblock \emph{Ann. of Statist.}, 28:\penalty0 393--400, 2000.

\bibitem[Schwarz(1978)]{sch78}
G.~Schwarz.
\newblock Estimating the dimension of a model.
\newblock \emph{Annals of statistics}, 6:\penalty0 461--464, 1978.

\bibitem[Simonoff(1996)]{simonoff1996}
J.~S. Simonoff.
\newblock \emph{Smoothing Methods in Statistics}.
\newblock Springer, New York, 1996.

\bibitem[Tukey(1977)]{tukey1977}
J.~W. Tukey.
\newblock \emph{Exploratory Data Analysis}.
\newblock Addison-Wesley, 1977.

\bibitem[Wendelberger(1982)]{wendelberger1982}
J.~Wendelberger.
\newblock Smoothing noisy data with multivariate splines and generalized
  cross-validation.
\newblock \emph{Ph.D thesis, University of Wisconsin}, 1982.

\bibitem[Wood(2011)]{mgcv}
S.~N. Wood.
\newblock \emph{\pkg{mgcv}: GAMs with GCV/AIC/REML smoothness estimation and
  GAMMs by PQL}, 2011.
\newblock URL \url{http://CRAN.R-project.org/package=mgcv}.
\newblock R package version 1.7.5.

\end{thebibliography}
\end{document}